\documentclass{ptephy_v1}

\usepackage{amsmath}
\usepackage{hyperref} 
\usepackage{aas_macros4ptep}
\usepackage{braket}

\begin{document}

\title{Role of symmetry energy at subnuclear densities in protoneutron star crusts}


\author[1,*]{Ken'ichiro Nakazato}
\affil[1]{Faculty of Arts and Science, Kyushu University, Fukuoka 819-0395, Japan
\email{nakazato@artsci.kyushu-u.ac.jp}}

\author[2,*]{Hajime Togashi}
\affil[2]{Department of Physics, Kyoto University, Kyoto 606-8502, Japan
\email{togashi.hajime.4r@kyoto-u.ac.jp}}

\author[3]{Kohsuke Sumiyoshi}
\affil[3]{National Institute of Technology, Numazu College, Shizuoka 410-8501, Japan}

\author[4]{Hideyuki Suzuki}
\affil[4]{Faculty of Science \& Technology, Tokyo University of Science, 2641 Yamazaki, Noda, Chiba 278-8510, Japan}



\begin{abstract}%
The impact of matter properties at subnuclear densities on the evolution of protoneutron stars is investigated. Several models of nuclear equation of state (EOS) are constructed with varying saturation parameters, particularly the symmetry energy $S_0$ and its density slope $L$. Using the Thomas--Fermi approximation, the mass and proton numbers of heavy nuclei at subnuclear densities are systematically evaluated, along with their dependence on the EOS. Cooling simulations of protoneutron stars reveal that EOSs with smaller $L$ values lead to a longer cooling timescale and higher average neutrino energies. This behavior is attributed to the enhanced neutrino scattering caused by larger mass numbers, which increases the thermal insulation. Furthermore, the crystallization temperature, marking the onset of crust formation, is found to be higher for EOSs with smaller values of $L$. This is due to the enhanced Coulomb energy associated with larger proton numbers. As a result, despite slower cooling, crust formation occurs earlier for smaller-$L$ EOSs. These findings indicate that the timing of crust formation is sensitive to the EOS and highlight the importance of late-time neutrino observations as probes of the matter properties at subnuclear densities.
\end{abstract}

\subjectindex{D41, D42, E26, E30, E32, F22}

\maketitle

\section{Introduction} \label{sec:intro}

Neutron stars are formed as remnants of core-collapse supernova explosions. They are initially born as hot protoneutron stars and cool over tens of seconds via neutrino emission~\cite{1975PThPh..53..595S}. Structurally, cold neutron stars consist of a dense central core surrounded by a solid crust. The crust is composed of heavy nuclei settled in a Coulomb lattice, embedded in a degenerate electron gas. In the inner crust, these nuclei coexist with a gas of unbound neutrons dripped out of the nuclei~\cite{1971NuPhA.175..225B,1971ApJ...170..299B}. In protoneutron stars with high temperatures, however, the thermal energy prevents heavy nuclei from forming a Coulomb lattice, and hence the solid crust is not expected to exist. The formation of the crust, which occurs as a result of the thermal structure evolution of neutron stars, is phenomenologically important~\cite{2014PASJ...66L...1S}. For instance, since a solid crust is thought to play a role in sustaining the magnetic field generated during core collapse, crust formation would be crucial for understanding the origin and evolution of neutron star magnetic fields~\cite{2008AIPC..983..391S}. Not only the cooling processes by neutrinos but also the properties of nuclear matter are responsible for the timescale from protoneutron star formation to crust crystallization.

The macroscopic nuclear model developed by Oyamatsu and Iida~\cite{2003PThPh.109..631O} provided a systematic and comprehensive framework for connecting the gross features of atomic nuclei to the properties of matter at subnuclear densities and neutron star crusts~\cite{2007PhRvC..75a5801O}. By parameterizing the energy density of uniform nuclear matter and incorporating the phenomenological inhomogeneity energy, they obtained hundreds of sets of interaction models that achieve good agreement with the empirical mass and radius data of stable nuclei. This framework linked interaction models with the saturation parameters of nuclear matter, such as incompressibility $K$ and the density slope of the symmetry energy $L$, and enabled the consistent construction of the equation of state (EOS) that spans a broad range of $K$ and $L$. Furthermore, the mass numbers $A$ and proton numbers $Z$ of nuclei and the critical density for the transition to uniform matter in the crust of cold neutron stars were predicted systematically. Importantly, these characteristic features of inhomogeneous nuclear matter were shown to be sensitive to $L$, as described below.

In this work, using interaction parameters updated recently~\cite{2023PTEP.2023f3D03O}, we extend the model by Oyamatsu and Iida to finite temperatures. Originally, their model was developed and applied at zero temperature. However, the evolution of protoneutron stars requires an EOS applicable to finite temperatures, especially in regions spanning the inhomogeneous and uniform phases. By incorporating thermal effects into both phases, we construct a temperature-dependent EOS that captures the essential physics of hot nuclear matter, including the characteristic features of the inhomogeneous phase. Then, we extend the EOS models to supranuclear densities and apply them to simulations of protoneutron star cooling so as to explore the role of the $L$ parameter in neutrino emission and crust crystallization.

Generally, EOSs with smaller values of $L$ are expected to result in slower cooling of protoneutron stars. This is attributable to two effects: first, smaller-$L$ EOSs tend to predict more compact neutron stars with smaller radii~\cite{2014PTEP.2014e1E01S,2020PTEP.2020d3D01H}, which increases the Kelvin--Helmholtz timescale~\cite{2020ApJ...891..156N}; second, such EOSs yield larger mass numbers $A$~\cite{2007PhRvC..75a5801O}, which enhances neutrino-nucleus coherent scattering and thus increases thermal insulation~\cite{2018PhRvC..97c5804N}. On the other hand, the crystallization temperature of the crust is expected to be higher for smaller values of $L$, owing to the fact that the corresponding proton numbers $Z$ are systematically larger~\cite{2007PhRvC..75a5801O}, and the crystallization temperature is known to increase with $Z$~\cite{1990ApJ...361..511O,2007PhRvE..75f6101H,2008LRR....11...10C}. These competing effects raise a key question: how does the time required for crust crystallization depend on $L$? In particular, it is important to determine whether slower cooling or earlier crystallization has a dominant impact. In this work, we address this question by applying the finite-temperature extension of the macroscopic nuclear model to protoneutron star evolution and systematically analyzing the $L$-dependence of the crust formation.

This paper is organized as follows. In \S~\ref{sec:eos}, we describe the construction of the EOS models discussed in this study. We also introduce the crystallization condition and examine the $L$ dependence of the crust formation region in the phase diagram. In addition, we show the mass--radius relations of neutron stars, obtained by extending the EOS to high densities. In \S~\ref{sec:result}, we present the results of protoneutron star cooling simulations. In particular, we focus on the temporal evolution of matter and crystallization temperature profiles to determine the onset of crust formation in protoneutron stars. Finally, we devote \S~\ref{sec:summary} to conclusions and discussion.

\section{Nuclear EOS models}
\label{sec:eos}

To describe matter at low densities and temperatures, where nuclear inhomogeneities are energetically favored, we construct the EOS for nuclear matter by employing a finite-temperature Thomas--Fermi approximation. This approach is based on the framework developed in Refs.~\cite{2011ApJS..197...20S, 2017NuPhA.961...78T}, which extends the original zero-temperature Thomas--Fermi method formulated by Oyamatsu and Iida~\cite{2003PThPh.109..631O, 2007PhRvC..75a5801O, 2023PTEP.2023f3D03O} to finite temperatures.

In this model, matter in a Wigner--Seitz (WS) cell is composed of free nucleons, alpha particles, and a representative heavy nucleus.  
The local number densities of protons and neutrons, \(n_{\mathrm{p}}(r)\) and \(n_{\mathrm{n}}(r)\), at position $\boldsymbol{r}$ inside the cell include contributions from both free nucleons and nucleons bound inside the heavy nucleus.  
Alpha particles are treated as a separate species, with local number density \(n_\alpha(r)\).

With these definitions, the nucleon number density averaged over the WS cell is defined by
\begin{equation}
n_{\mathrm{B}} = \frac{1}{V_{\mathrm{cell}}} \int_{\mathrm{cell}} d\mbox{\boldmath$r$} \, \left[ n_{\mathrm{p}}(r)+n_{\mathrm{n}}(r)+4n_{\alpha}(r) \right],
\label{nBnonuni}
\end{equation}
where \( V_{\mathrm{cell}} \) is the volume of the WS cell. 
Thus, although the presence of the heavy nucleus is not explicitly visible in Eq.~\eqref{nBnonuni}, the integration over the WS cell properly accounts for all nucleons in the system.
The average proton fraction within the cell is defined by
\begin{equation}
Y_{\mathrm{p}} = \frac{1}{n_{\mathrm{B}}V_{\mathrm{cell}}} \int_{\mathrm{cell}} d\mbox{\boldmath$r$} \, \left[ n_{\mathrm{p}}(r)+2n_{\alpha}(r) \right].
\label{Ypnonuni}
\end{equation} To account for the finite volume occupied by alpha particles, we introduce an excluded volume correction, such that the available volume fraction for nucleons and alpha particles is given by \((1 - u)\), where \(u = n_\alpha (r) v_\alpha\), with \(v_\alpha=24~\mathrm{fm}^3\) denoting the volume of a single alpha particle. The total free energy in the WS cell at temperature $T$ is then expressed as  
\begin{equation}  
F_{\mathrm{cell}} = F_{\mathrm{bulk}} + E_{\mathrm{grad}} + E_{\mathrm{C}},  
\end{equation}  
where the bulk part is written as  
\begin{equation}  
F_{\mathrm{bulk}} = \int_{\mathrm{cell}} d\boldsymbol{r}~(1 - u) \left[ f_{\mathrm{N}}(\tilde{n}_{\mathrm{p}} (r), \tilde{n}_{\mathrm{n}}(r), T) + f_{\alpha}(\tilde{n}_\alpha(r), T) \right].  
\end{equation}  
Here, \(\tilde{n}_{\mathrm{p}}(r)\), \(\tilde{n}_{\mathrm{n}}(r)\), and \(\tilde{n}_\alpha(r)\) are the local number densities of protons, neutrons, and alpha particles normalized by the available volume \((1 - u)\), and \(f_{\mathrm{N}}\) and \(f_{\alpha}\) are the free energy densities of uniform nucleonic and alpha matter, respectively. 

The free energy density of uniform nuclear matter is expressed as
\begin{equation}
f_{\mathrm{N}}(n_{\mathrm{p}}, n_{\mathrm{n}}, T) = f_{\mathrm{F}}(n_{\mathrm{p}}, n_{\mathrm{n}}, T) 
+\left(1-\delta^2 \right) \left[a_1 n^2 +\frac{a_2 n^3}{1+a_3 n}\right] +\delta^2 \left[b_1 n^2 +\frac{b_2 n^3}{1+b_3 n}\right],
\label{eq:funi}
\end{equation}
where \(n = n_{\mathrm{n}} + n_{\mathrm{p}}\) and \(\delta = (n_{\mathrm{n}} - n_{\mathrm{p}})/n\) represent the total nucleon number density and neutron-proton asymmetry, respectively. Note that if the matter is in a uniform phase, the nucleon number density $n$ coincides with the average baryon number density $n_{\mathrm{B}}$. The term \(f_{\mathrm{F}}\) denotes the free energy density of a non-interacting Fermi gas. At zero temperature, this expression reproduces the kinetic energy density used in the original Thomas--Fermi model by Oyamatsu and Iida. The coefficients \(a_1\)–\(a_3\) and \(b_1\)–\(b_3\) characterize the potential energy density of symmetric and asymmetric nuclear matter, respectively.
In this model, the temperature dependence of the EOS originates solely from the kinetic term in \(f_{\mathrm{F}}\), as the potential energy remains temperature independent. This treatment is in line with the behavior of the Skyrme-type density functional theory~\cite{2025Symm...17..445T}.  
The contribution from alpha particles, \(f_\alpha\), follows the prescription given in Eq.~(8) of Ref.~\cite{2017NuPhA.961...78T}.

The gradient part is given by  
\begin{equation}  
E_{\mathrm{grad}} = F_0 \int_{\mathrm{cell}} d\boldsymbol{r}~ \left| \nabla \left(n_{\mathrm{p}}(r) + n_{\mathrm{n}}(r)\right) \right|^2,    
\end{equation}  
where the coefficient \(F_0\) is the inhomogeneity parameter. 
The six parameters \( a_1 \)–\( a_3 \) and \( b_1 \)–\( b_3 \) in Eq.~(3), along with \( F_0 \), are calibrated to reproduce bulk properties of stable nuclei, such as the neutron excess, mass excess, and root-mean-square charge radius.
In this study, as shown later, we adopt four parameter sets out of the 304 optimized sets presented in Ref.~\cite{2023PTEP.2023f3D03O}.

The Coulomb energy part is expressed as  
\begin{equation}  
E_{\mathrm{C}} = \frac{e^2}{2} \int_{\mathrm{cell}} d\boldsymbol{r} \int_{\mathrm{cell}} d\boldsymbol{r}' \frac{n_{\mathrm{ch}}(r) n_{\mathrm{ch}}(r')}{|\boldsymbol{r} - \boldsymbol{r}'|} + c_{\mathrm{bcc}} \frac{(Z_{\mathrm{non}} e)^2}{a}, 
\label{eq:ec}
\end{equation}  
with the charge density defined as  
\begin{equation}  
n_{\mathrm{ch}}(r) = n_{\mathrm{p}}(r) + 2 n_\alpha(r) - n_e,  
\end{equation}  
where \(n_e = Y_{\mathrm{p}} n_{\mathrm{B}}\) ensures net charge neutrality.
In Eq.~(\ref{eq:ec}), \(a\) is the lattice constant defined by $V_{\mathrm{cell}} = a^3$. The parameter \(Z_{\mathrm{non}}\) represents the non-uniform part of the total charge number per cell, and is defined as in Eq.~(21) of Ref.~\cite{2017NuPhA.961...78T}. The numerical factor \(c_{\mathrm{bcc}} = 0.006562\) corresponds to the Coulomb energy coefficient for a body-centered cubic (bcc) lattice~\cite{1993NuPhA.561..431O}.

We adopt parametrized forms for the density profiles of each component, ${n}_{\mathrm{p}}(r)$, ${n}_{\mathrm{n}}(r)$, and $n_\alpha(r)$, and numerically minimize the free energy density $F_{\mathrm{cell}}/V_{\mathrm{cell}}$ with respect to the profile parameters at fixed $n_{\mathrm{B}}$, $Y_{\mathrm{p}}$, and $T$. The final free energy per baryon is given by
\begin{equation}
F(n_{\mathrm{B}}, Y_{\mathrm{p}}, T) = \frac{1}{n_{\mathrm{B}}} \left( \frac{F_{\mathrm{cell}}}{V_{\mathrm{cell}}} \right).
\end{equation}
By comparing this free energy to that of uniform matter at the same thermodynamic point, we determine whether inhomogeneous phases are favored. Once the equilibrium phase is chosen, thermodynamic quantities are computed by differentiating $F(n_{\mathrm{B}}, Y_{\mathrm{p}}, T)$ accordingly.

The nuclear saturation parameters used in this study are summarized in Table~\ref{tab:parameters}. These EOS models are taken from Ref.~\cite{2023PTEP.2023f3D03O}, and correspond to different choices of the symmetry energy slope parameter \(L\), while keeping the incompressibility \(K\) fixed at 230~MeV. The saturation density $n_0$, the saturation energy $w_0$, and the symmetry energy $S_0$ are optimal values. The labels of the EOS models, except for EOS Z, follow those used in Ref.~\cite{2007PhRvC..75a5801O}.

\begin{table}[t]
    \caption{EOS models from Ref.~\cite{2023PTEP.2023f3D03O}.}
    \centering
    \begin{tabular}{cccccc}
    \hline
    EOS & $n_0$ (MeV) & $w_0$ (MeV) & $K$ (MeV) & $S_0$ (MeV) & $L$ (MeV) \\
    \hline
    B & 0.160 & $-16.2$ & 230 & 33.6 & 73.4 \\
    E & 0.160 & $-16.2$ & 230 & 31.1 & 42.6 \\
    Z & 0.160 & $-16.1$ & 230 & 29.7 & 23.7 \\
    H & 0.160 & $-16.1$ & 230 & 28.5 & 7.6  \\
    \hline
    \end{tabular}
    \label{tab:parameters}
\end{table}

In the Thomas--Fermi model described above, the free energy is evaluated under the assumption that the nuclei are crystallized and behave as ions in the solid state. Nonetheless, whether this is actually the case depends on the Coulomb coupling parameter, which is written as \cite{1990ApJ...361..511O}
\begin{equation}
\Gamma=\frac{(Ze)^2}{Rk_{\rm B}T},
\label{eq:gamma}
\end{equation}
where $Z$ is the charge number of an ion in the WS cell, which is defined as in Eq.~(25) of Ref.~\cite{2017NuPhA.961...78T}, and $k_{\rm B}$ is the Boltzmann constant. The ion sphere radius $R$ is related to other physical quantities as
\begin{equation}
R=\left(\frac{3V_{\rm cell}}{4\pi}\right)^{1/3}=\left(\frac{3A^\prime}{4\pi n_{\rm B}}\right)^{1/3}=\left(\frac{3A}{4\pi X_{\rm A} n_{\rm B}}\right)^{1/3},
\label{eq:ion-radius}
\end{equation}
where $A^\prime$ is the number of nucleons per ion (sum of the mass number $A$ and the number of dripped nucleons per ion) and $X_{\rm A}$ is the fraction of ions. Since electrons are ultrarelativistic and strongly degenerate, they act as a uniform background of negative charge that neutralizes the average space charge of positive ions. Therefore, inhomogeneous nuclear matter can reasonably be modeled as a one-component plasma, whose crystallization to a bcc lattice occurs for $\Gamma>175$ \cite{2008LRR....11...10C}. Thus, the crystallization temperature of inhomogeneous nuclear matter is estimated as \begin{equation}
T_{\rm c} \approx 6.4\times10^9~{\rm K}~\left(\frac{\Gamma}{175}\right)^{-1}\left(\frac{\rho_{\rm B}}{10^{14}~{\rm g}~{\rm cm}^{-3}}\right)^{1/3}\left(\frac{Z}{26}\right)^{2}\left(\frac{A^\prime}{260}\right)^{-1/3},
\label{eq:crit-temp}
\end{equation}
where $\rho_{\rm B}$ is the baryon mass density. Strictly speaking, this expression ignores the effect of proton drip, which would reduce the Coulomb screening provided by electrons. However, since the evaluated amount of dripped protons is negligible, we adopt the expression above for simplicity.

\begin{figure}[t]
  \centering
  \begin{minipage}[t]{0.495\textwidth}
    \centering
    \includegraphics[width=\linewidth]{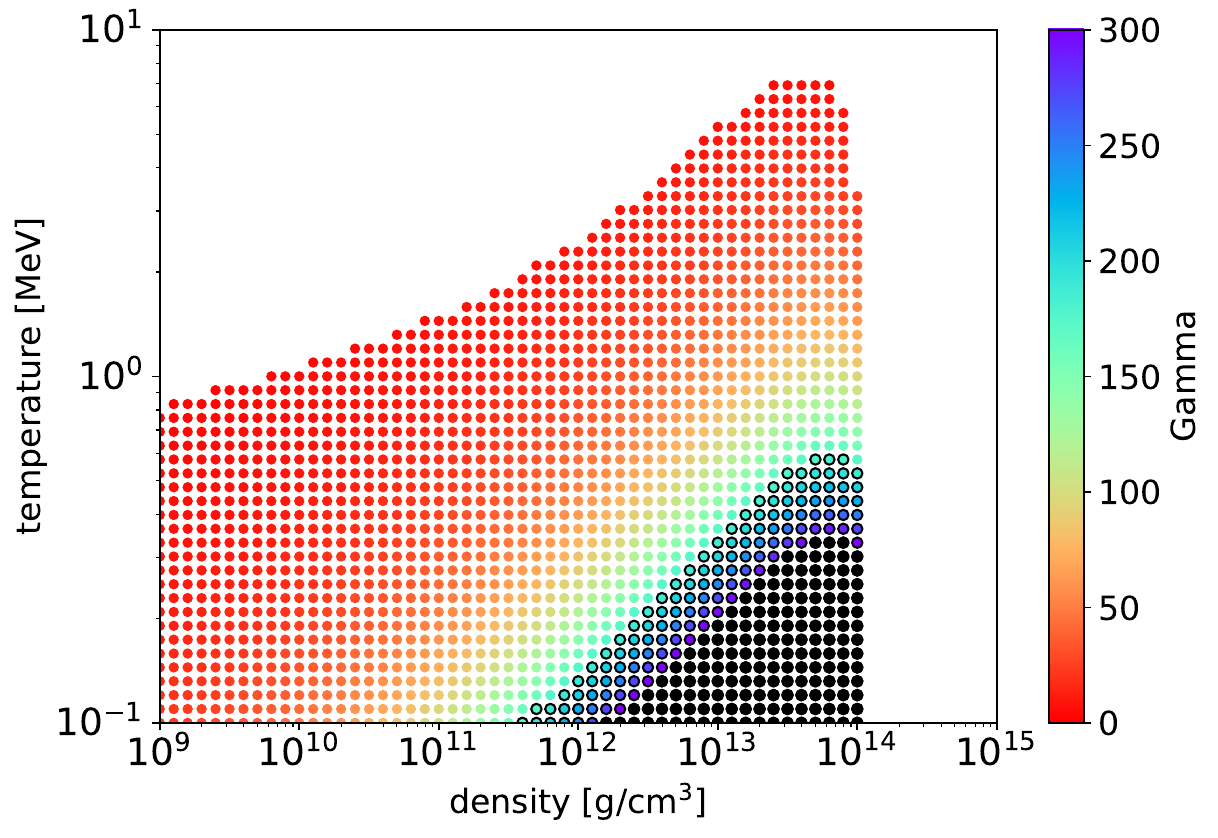}
    \vspace{0.5ex}
    {\small (a) EOS B ($L=73.4$~MeV, $K=230$~MeV)}
  \end{minipage}
  \hfill
  \begin{minipage}[t]{0.495\textwidth}
    \centering
    \includegraphics[width=\linewidth]{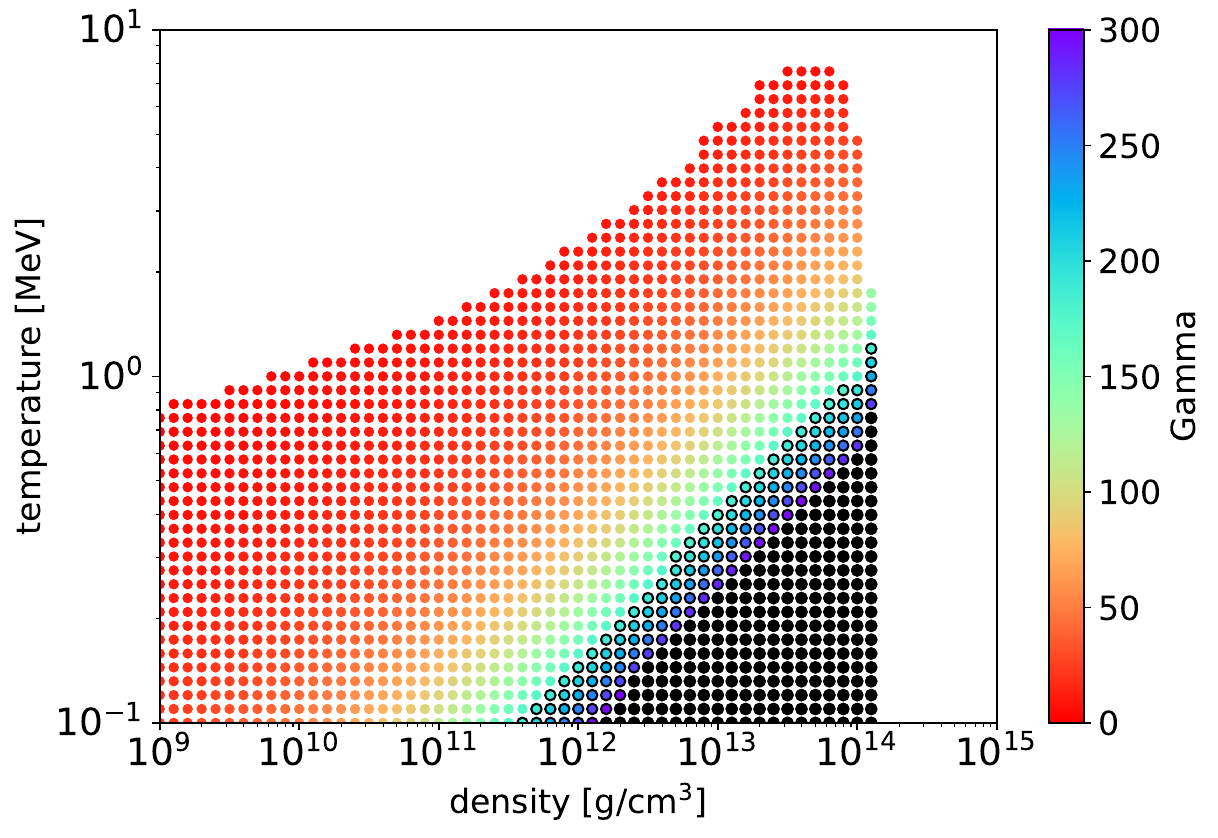}
    \vspace{0.5ex}
    {\small (b) EOS E ($L=42.6$~MeV, $K=230$~MeV)}
  \end{minipage}
  \vspace{1em}
  \begin{minipage}[t]{0.495\textwidth}
    \centering
    \includegraphics[width=\linewidth]{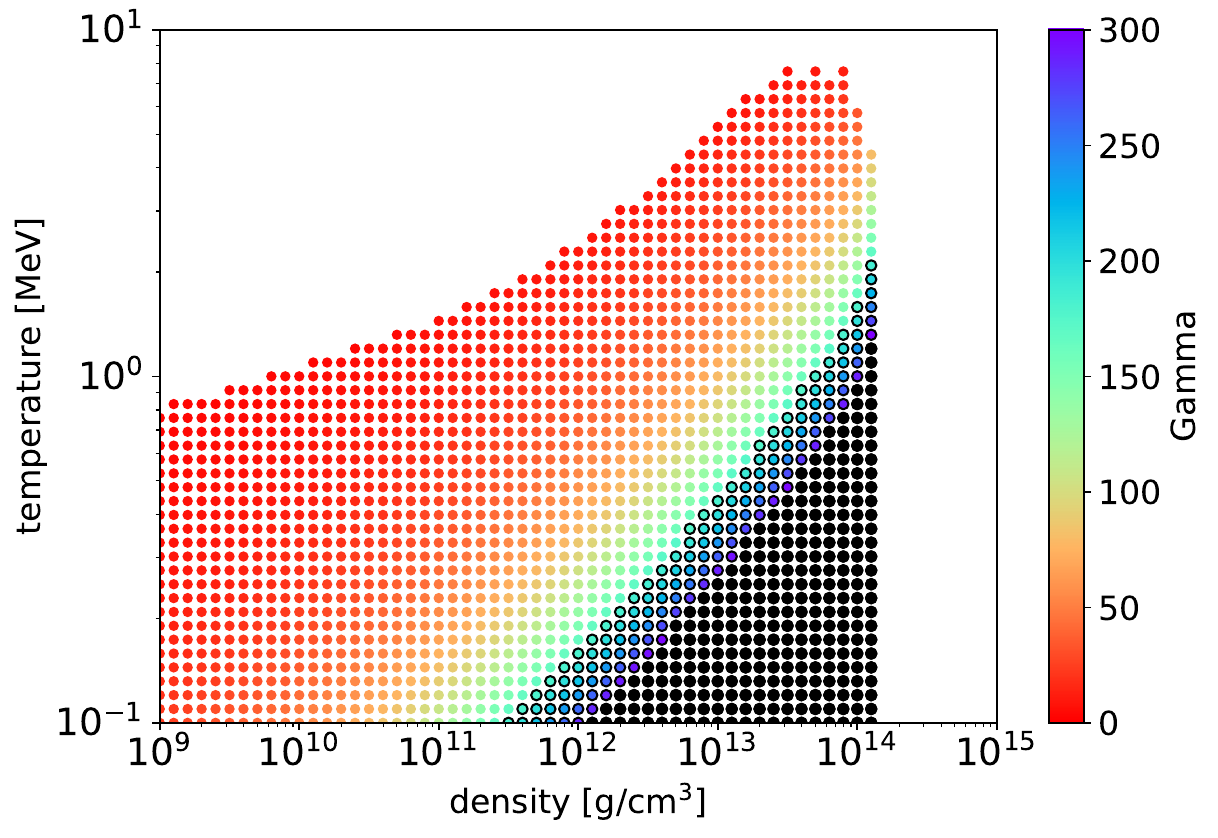}
    \vspace{0.5ex}
    {\small (c) EOS Z ($L=23.7$~MeV, $K=230$~MeV)}
  \end{minipage}
  \hfill
  \begin{minipage}[t]{0.495\textwidth}
    \centering
    \includegraphics[width=\linewidth]{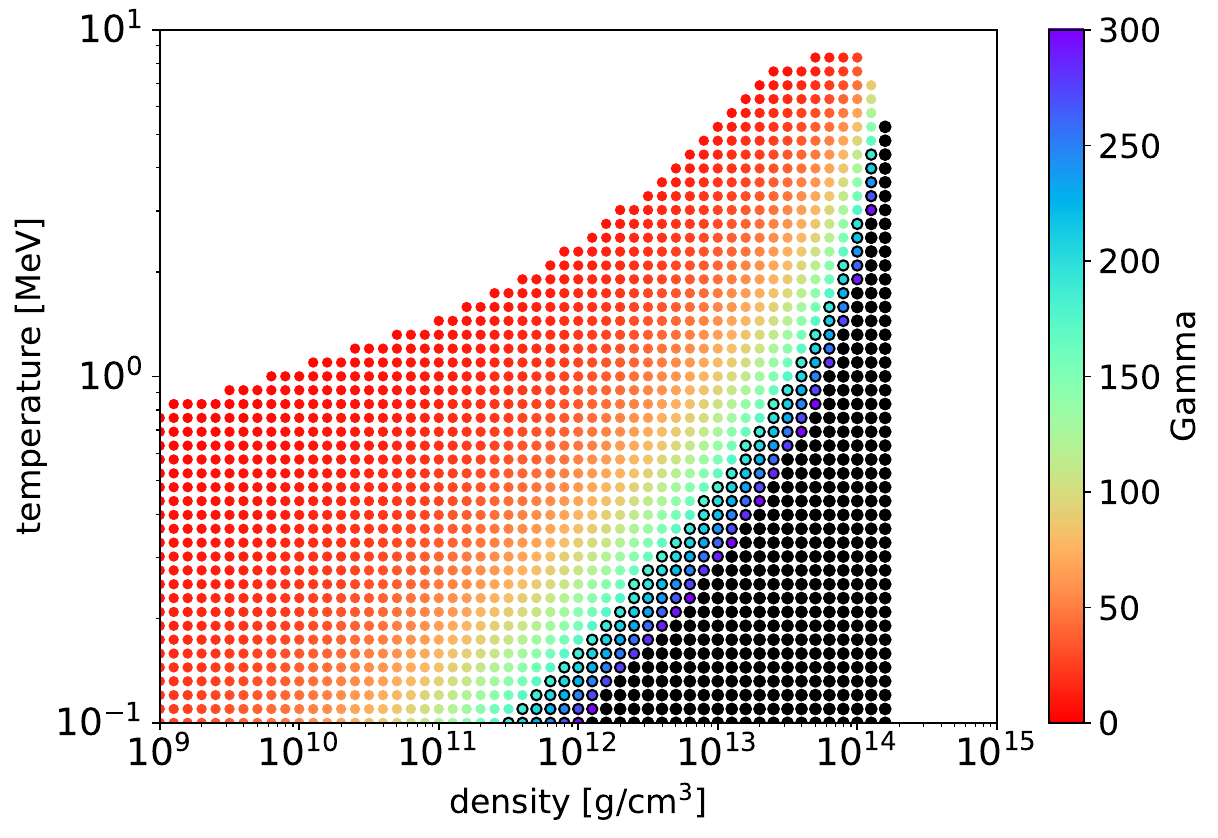}
    \vspace{0.5ex}
    {\small (d) EOS H ($L=7.6$~MeV, $K=230$~MeV)}
  \end{minipage}
  \caption{Phase diagrams of matter with $Y_{\rm p} = 0.04$ for (a) EOS B, (b) EOS E, (c) EOS Z, and (d) EOS H. The plotted regions represent the inhomogeneous phase, with the color scale indicating the value of $\Gamma$. Regions with fully black plots denote $\Gamma > 300$, and regions with black-outlined plots denote $\Gamma > 175$.}
  \label{fig:gamma_phase}
\end{figure}

In Fig.~\ref{fig:gamma_phase}, contour plots of $\Gamma$ are presented on the phase diagrams in the $\rho_{\rm B}$-$T$ plane for matter with $Y_{\rm p}=0.04$, based on the EOS models constructed as described above. Note that at subnuclear densities, the condition for neutrino-less $\beta$ equilibrium is typically satisfied when $Y_{\rm p}$ is approximately 0.04. The crystallization temperature as a function of density is identified by the threshold $\Gamma = 175$ in this figure. As already mentioned, the crystallization temperature is higher for EOS models with smaller values of $L$ because the corresponding proton numbers $Z$ are systematically larger. At subnuclear densities, smaller-$L$ EOSs have higher symmetry energies and, hence, higher surface tension, allowing nuclei to stabilize with higher proton numbers $Z$ \cite{2007PhRvC..75a5801O}. Furthermore, the transition density between the inhomogeneous and uniform phases is higher for smaller-$L$ EOSs (see Fig.~\ref{fig:gamma_phase}) because the onset density of proton clustering in uniform nuclear matter is higher \cite{2007PhRvC..75a5801O}. Therefore, crystallization in inhomogeneous nuclear matter occurs at higher temperatures for EOS models with smaller values of $L$.

The macroscopic nuclear model and the associated Thomas--Fermi approximation discussed so far are only valid below saturation density. However, regions with higher densities exist inside neutron stars and protoneutron stars. Therefore, it is necessary to consider EOS models that are extended to supranuclear densities. For this purpose, we adopt a simple parametric model proposed in Ref.~\cite{2019ApJ...878...25N}, where the energy per baryon at zero temperature is expressed as
\begin{equation}
w(n_{\rm B},Y_{\rm p}) = w_0 + \frac{K}{18n_0^2} (n_{\rm B}-n_0)^2 + S(n_{\rm B})\,(1-2Y_{\rm p})^2,
\label{eq:paraeos}
\end{equation}
for $n_{\rm B} > n_0$, and the symmetry energy $S(n_{\rm B})$, defined as the energy difference between symmetric nuclear matter ($Y_{\rm p}=0.5$) and pure neutron matter ($Y_{\rm p}=0$), is given by
\begin{equation}
S(n_{\rm B}) = S_0 + \frac{L}{3n_0} (n_{\rm B}-n_0) + \frac{1}{n_0^2} \left( S_{00} - S_0 -\frac{L}{3} \right)(n_{\rm B}-n_0)^2,
\label{eq:parasym}
\end{equation}
for $n_{\rm B} > n_0$. In the present study, each EOS listed in Table~\ref{tab:parameters} is extended to supranuclear densities using common saturation parameters, such as $S_0$ and $L$. Meanwhile, the parameter $S_{00}$, newly introduced in Eq. (\ref{eq:parasym}), represents the symmetry energy at a density of $2n_0$, which is typically assumed to be around 45~MeV \cite{2019EPJA...55...39Z,2020PhRvR...2b2033L,2020PhRvL.125t2702D}. In addition, EOS models with higher $S_0$ and larger $L$ tend to predict higher symmetry energy at supranuclear densities. Thus, as for the reference model, we choose $S_{00}=60$~MeV for EOS B ($L=73.4$~MeV), $S_{00}=50$~MeV for EOS E ($L=42.6$~MeV), $S_{00}=40$~MeV for EOS Z ($L=23.7$~MeV), and $S_{00}=30$~MeV for EOS H ($L=7.6$~MeV). So as to assess the effect of $S_{00}$, we also examine models with $S_{00}$ ranging from 40 to 60~MeV for EOS E and from 35 to 50~MeV for EOS Z. In the following, models are denoted as Xyy, where the letter X identifies the subnuclear-density EOS (B, E, Z or H), and yy indicates the assigned value of $S_{00}$. Incidentally, as already mentioned, the incompressibility is fixed at $K=230$~MeV for all EOS models. Furthermore, we extend the EOS at supranuclear densities to finite temperatures in a manner consistent with the subnuclear regime and write the free energy per baryon as
\begin{equation}
F(n_{\rm B},Y_{\rm p},T)= \frac{f_{\rm F}(Y_{\rm p}n_{\rm B},(1-Y_{\rm p})n_{\rm B},T)-f_{\rm F}(Y_{\rm p}n_{\rm B},(1-Y_{\rm p})n_{\rm B},0)}{n_{\rm B}}+ w(n_{\rm B},Y_{\rm p}),
\label{eq:suprafinite}
\end{equation}
where $f_{\rm F}$ denotes the same function as given in Eq.~(\ref{eq:funi}). Note that, since $w$ includes not only the potential contribution but also the kinetic contribution, the kinetic energy density at zero temperature, $f_{\rm F}(Y_{\rm p}n_{\rm B},(1-Y_{\rm p})n_{\rm B},0)$, is subtracted in Eq.~(\ref{eq:suprafinite}).

\begin{figure}[t]
\centering
  \begin{minipage}[t]{0.325\textwidth}
    \centering
    \includegraphics[width=\linewidth]{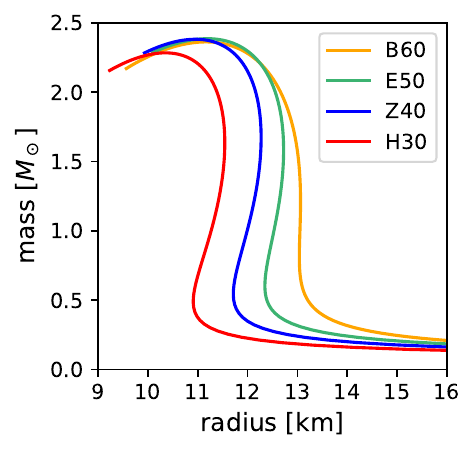}
    \vspace{0.5ex}
    {\small (a) Reference models}
  \end{minipage}
  \hfill
  \begin{minipage}[t]{0.325\textwidth}
    \centering
    \includegraphics[width=\linewidth]{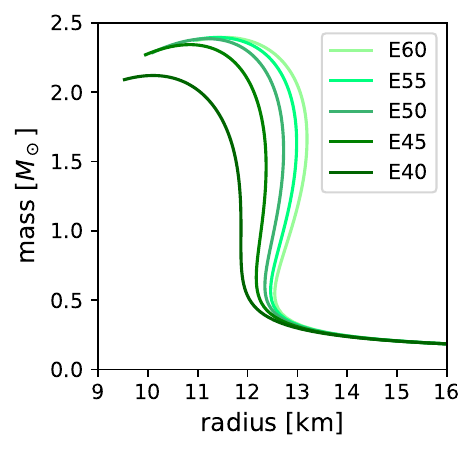}
    \vspace{0.5ex}
    {\small (b) Models of EOS E}
  \end{minipage}
  \hfill
  \begin{minipage}[t]{0.325\textwidth}
    \centering
    \includegraphics[width=\linewidth]{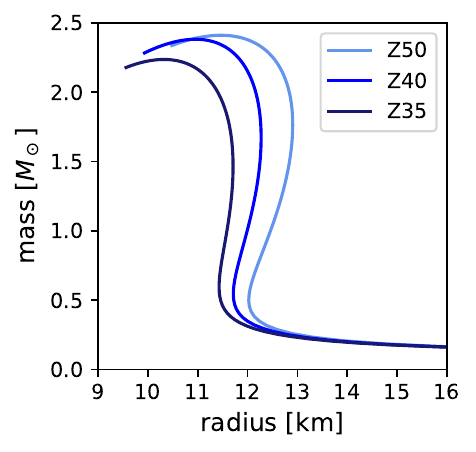}
    \vspace{0.5ex}
    {\small (c) Models of EOS Z}
  \end{minipage}
\caption{Mass-radius relations of cold neutron stars for (a) reference models, (b) models of EOS E, and (c) models of EOS Z. In (a), orange, green, blue, and red lines correspond to the models B60, E50, Z40, and H30, respectively. In (b) and (c), darker lines correspond to models with smaller values of $S_{00}$.}
\label{fig:MR}
\end{figure}

In Fig.~\ref{fig:MR}, we plot the mass--radius relations of neutron stars, assuming that the matter is composed of neutrons, protons, electrons and muons in $\beta$-equilibrium and satisfying charge neutrality. For low-mass neutron stars, EOSs with larger $L$ result in larger radii, while the neutron star radius is not significantly affected by $S_{00}$. In contrast, for canonical-mass neutron stars around $1.4M_\odot$, larger $S_{00}$ values lead to larger radii provided that $L$ remains unchanged. These trends indicate that larger $L$ and $S_{00}$, which correspond to higher symmetry energy at supranuclear densities, result in a less compressible EOS.

\section{Cooling of protoneutron stars and crust formation}
\label{sec:result}

We perform numerical simulations of protoneutron star cooling after core-collapse supernovae using a quasi-hydrostatic evolution code with approximate neutrino transport in general relativity under spherical symmetry \cite{1994pan..conf..763S,2019ApJ...878...25N}. The evolution is computed by solving the general relativistic hydrostatic equilibrium (Tolman--Oppenheimer--Volkoff) equation and the neutrino transport equation. Neutrino transfer is treated using the multigroup flux-limited diffusion scheme \cite{1994pan..conf..763S}, which governs the time evolution of the entropy and lepton-number profiles. We consider three species of neutrinos---$\nu_e$, $\bar\nu_e$, $\nu_x$ (with $\nu_x$ representing $\nu_\mu$, $\bar\nu_\mu$, $\nu_\tau$, and $\bar\nu_\tau$)---and adopt weak interaction rates primarily from Ref.~\cite{1985ApJS...58..771B}, supplemented with nucleon bremsstrahlung \cite{1993fna..conf..219S} and plasmon decay \cite{1986ApJ...310..815K}. Incidentally,  for simplicity, we do not include additional effects like mass accretion or convection.

The initial condition of the simulations is the same as in Ref.~\cite{2019ApJ...878...25N}, which is derived from hydrodynamic simulations of core collapse, bounce, and shock propagation, based on a general relativistic neutrino-radiation hydrodynamics code \cite{2005ApJ...629..922S} using a $15M_\odot$ progenitor model \cite{1995ApJS..101..181W} and the Togashi EOS \cite{2017NuPhA.961...78T}. We extract entropy and electron fraction profiles inside the stalled shock, which is located at the baryon mass coordinate of $1.47M_\odot$, and use it as the initial configuration of the protoneutron star. The final configuration of the protoneutron star models considered in this study is a cold neutron star with a baryon mass of $1.47M_\odot$, which corresponds to a gravitational mass of approximately $1.33M_\odot$, depending on the EOS.

\begin{figure}[t]
\centering
  \begin{minipage}[t]{0.325\textwidth}
    \centering
    \includegraphics[width=\linewidth]{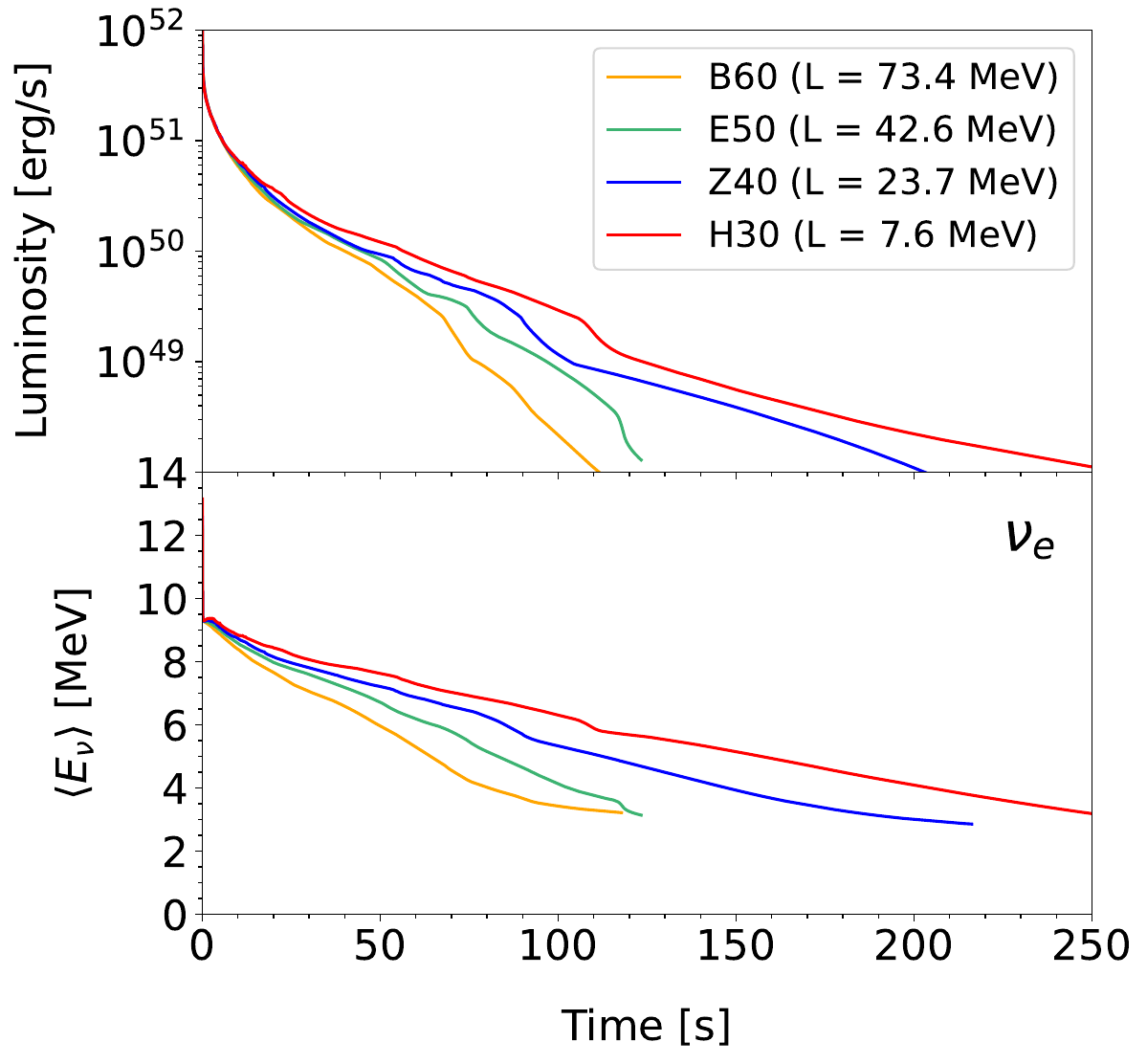}
    \vspace{0.5ex}
    {\small (a) $\nu_e$}
  \end{minipage}
  \hfill
  \begin{minipage}[t]{0.325\textwidth}
    \centering
    \includegraphics[width=\linewidth]{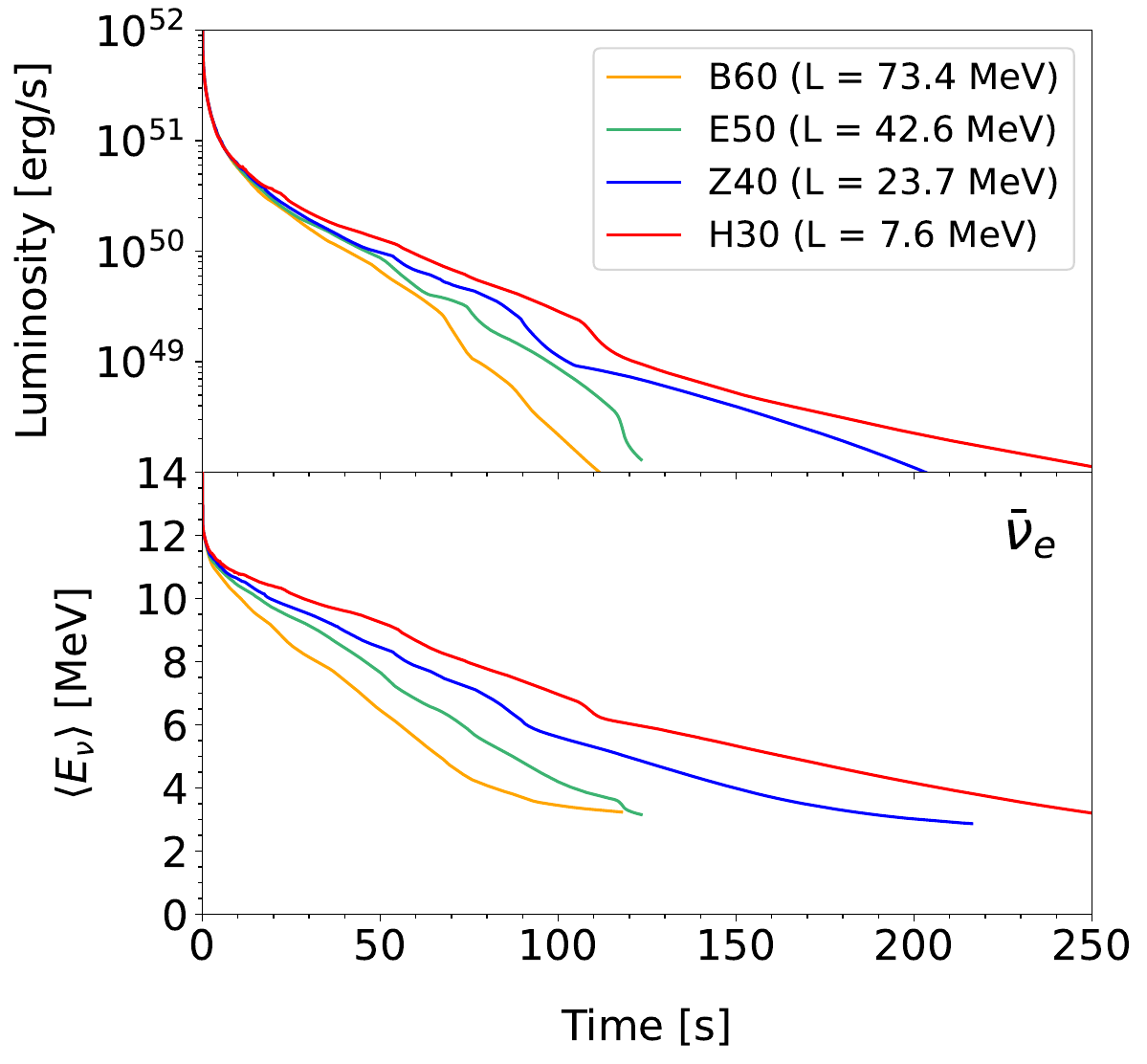}
    \vspace{0.5ex}
    {\small (b) $\bar\nu_e$}
  \end{minipage}
  \hfill
  \begin{minipage}[t]{0.325\textwidth}
    \centering
    \includegraphics[width=\linewidth]{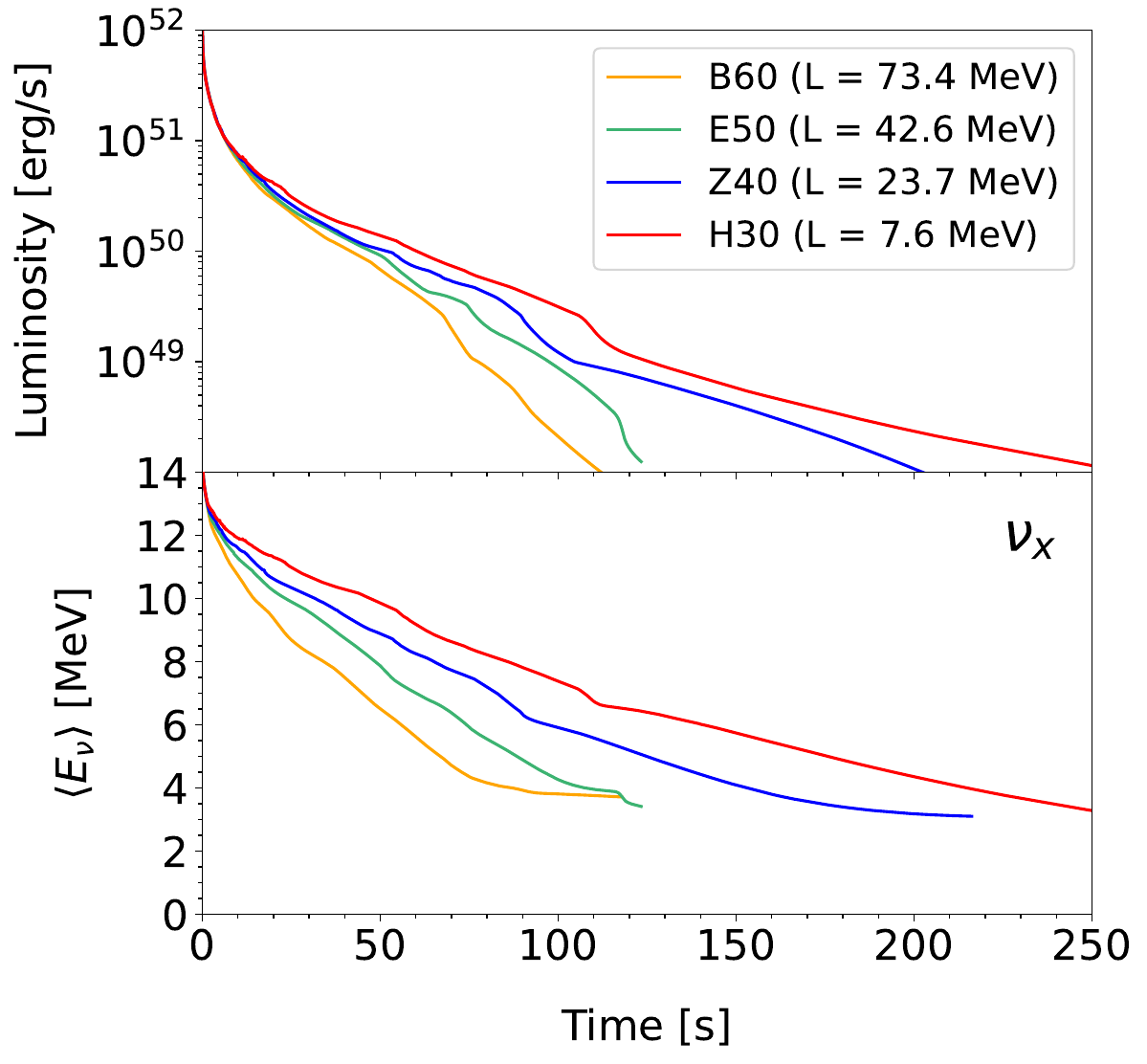}
    \vspace{0.5ex}
    {\small (c) $\nu_x$}
  \end{minipage}
\caption{Luminosity (upper) and average energy (lower) of neutrinos as a function of time for (a) $\nu_e$, (b) $\bar\nu_e$, and (c) $\nu_x$. The line colors follow the same definitions as in Fig.~\ref{fig:MR}(a).}
\label{fig:nulc}
\end{figure}

The cooling process of a protoneutron star can be observed as a time variation in the emitted neutrinos. In Fig.~\ref{fig:nulc}, we present the neutrino luminosity and average energy as functions of time for the reference models. We find that, in models with small $L$ (EOSs Z and H), neutrino emission persists significantly longer---even at low luminosities ($\lesssim10^{49}$~erg~s$^{-1}$)---compared to those with large $L$ (EOSs B and E). As already mentioned, EOSs with smaller $L$ exhibit a broader density range of the inhomogeneous phase. Furthermore, for smaller-$L$ EOSs, nuclei in the inhomogeneous phase have higher proton numbers $Z$ and consequently larger mass numbers $A$. These heavier nuclei have substantially larger scattering cross sections with neutrinos as a result of coherent effects, which significantly enhance neutrino--nucleus interactions. Increased scattering shortens the mean free path of neutrinos, effectively trapping neutrinos and prolonging their escape time \cite{2018PhRvC..97c5804N}.

\begin{figure}[t]
\centering
  \begin{minipage}[t]{0.325\textwidth}
    \centering
    \includegraphics[width=\linewidth]{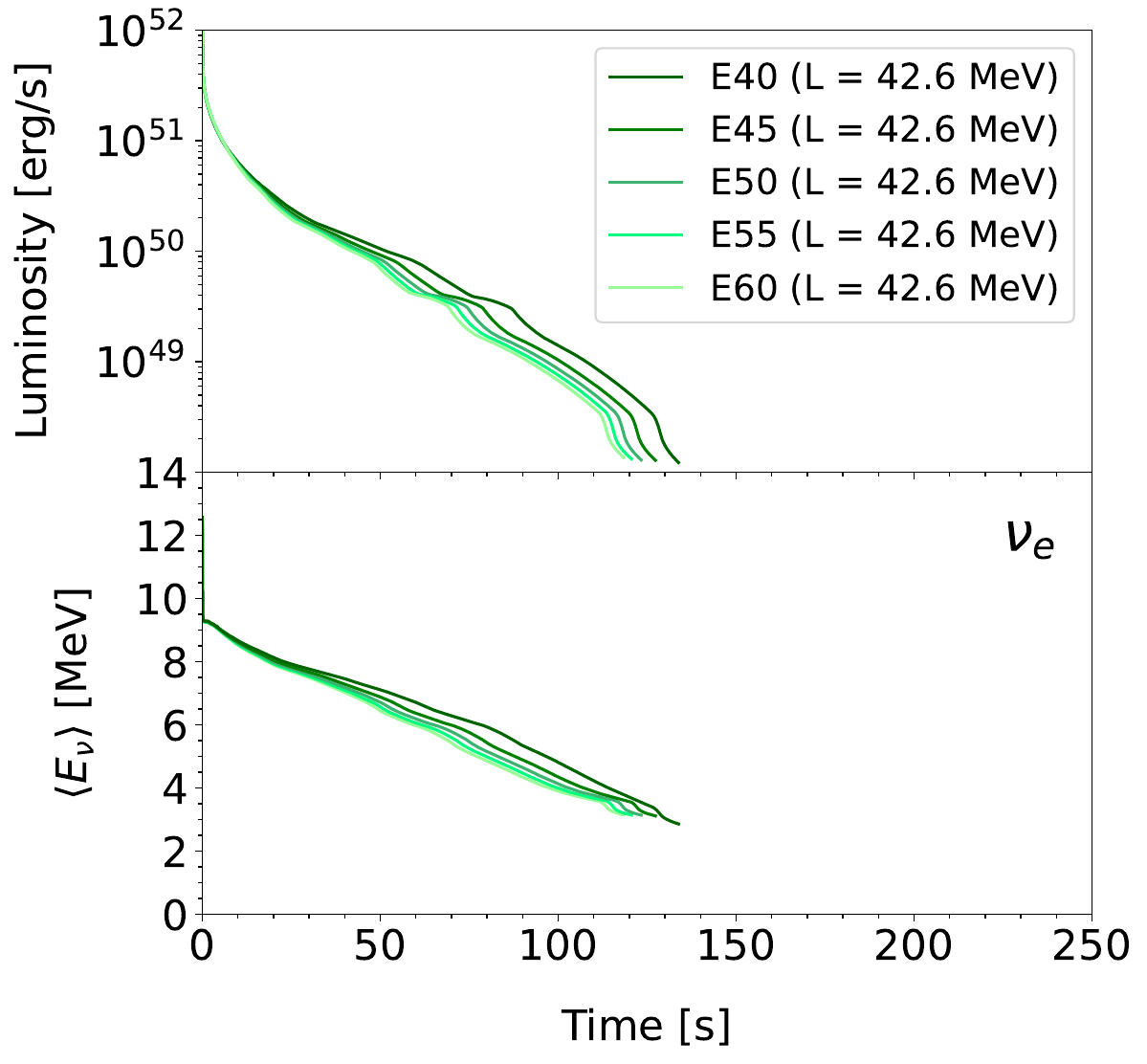}
    \vspace{0.5ex}
    {\small (a) $\nu_e$}
  \end{minipage}
  \hfill
  \begin{minipage}[t]{0.325\textwidth}
    \centering
    \includegraphics[width=\linewidth]{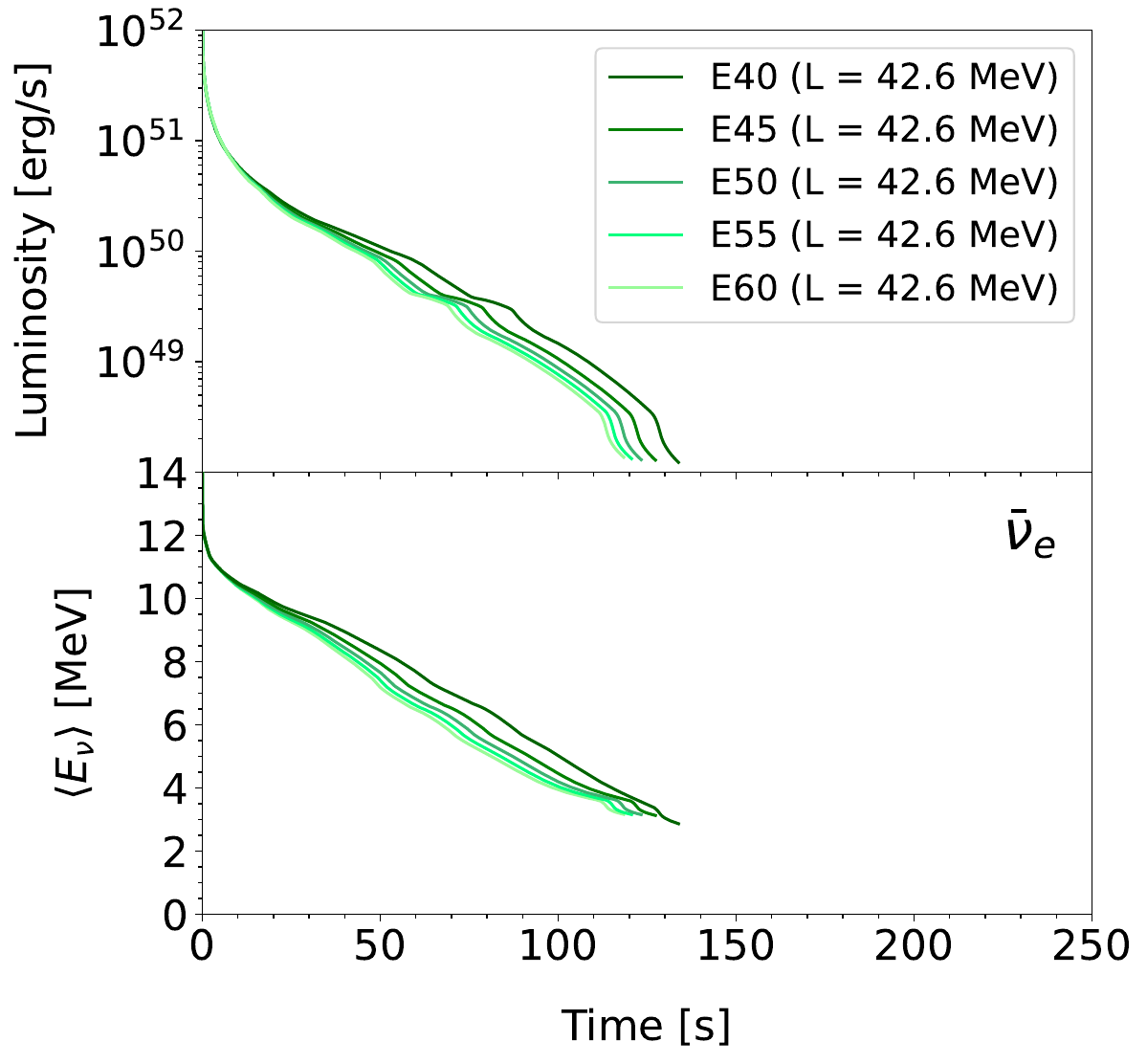}
    \vspace{0.5ex}
    {\small (b) $\bar\nu_e$}
  \end{minipage}
  \hfill
  \begin{minipage}[t]{0.325\textwidth}
    \centering
    \includegraphics[width=\linewidth]{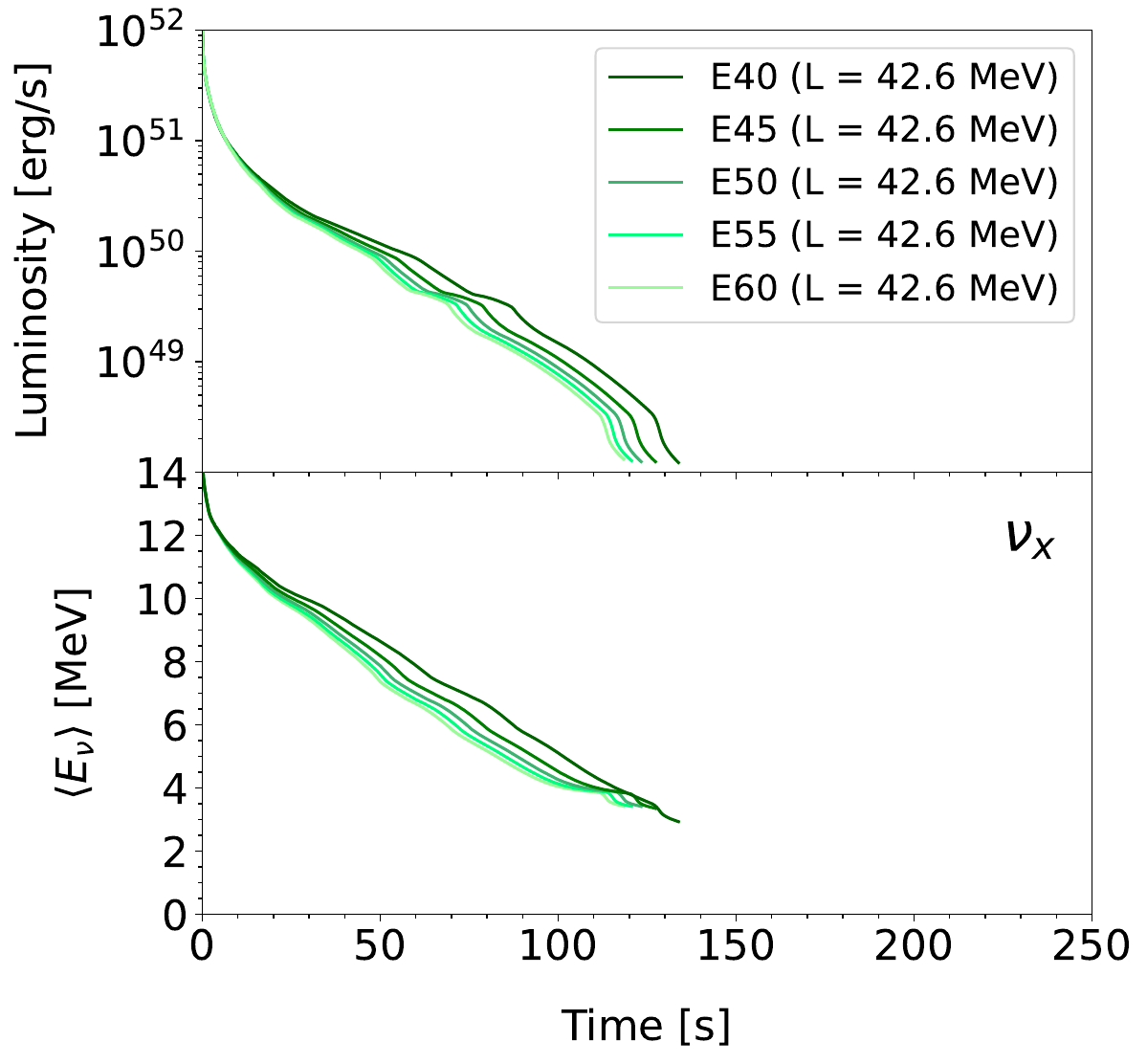}
    \vspace{0.5ex}
    {\small (c) $\nu_x$}
  \end{minipage}
\caption{Same as Fig.~\ref{fig:nulc} but for the models of EOS~E. The line colors follow the same definitions as in Fig.~\ref{fig:MR}(b).}
\label{fig:nulcE}
\end{figure}

\begin{figure}[t]
\centering
  \begin{minipage}[t]{0.325\textwidth}
    \centering
    \includegraphics[width=\linewidth]{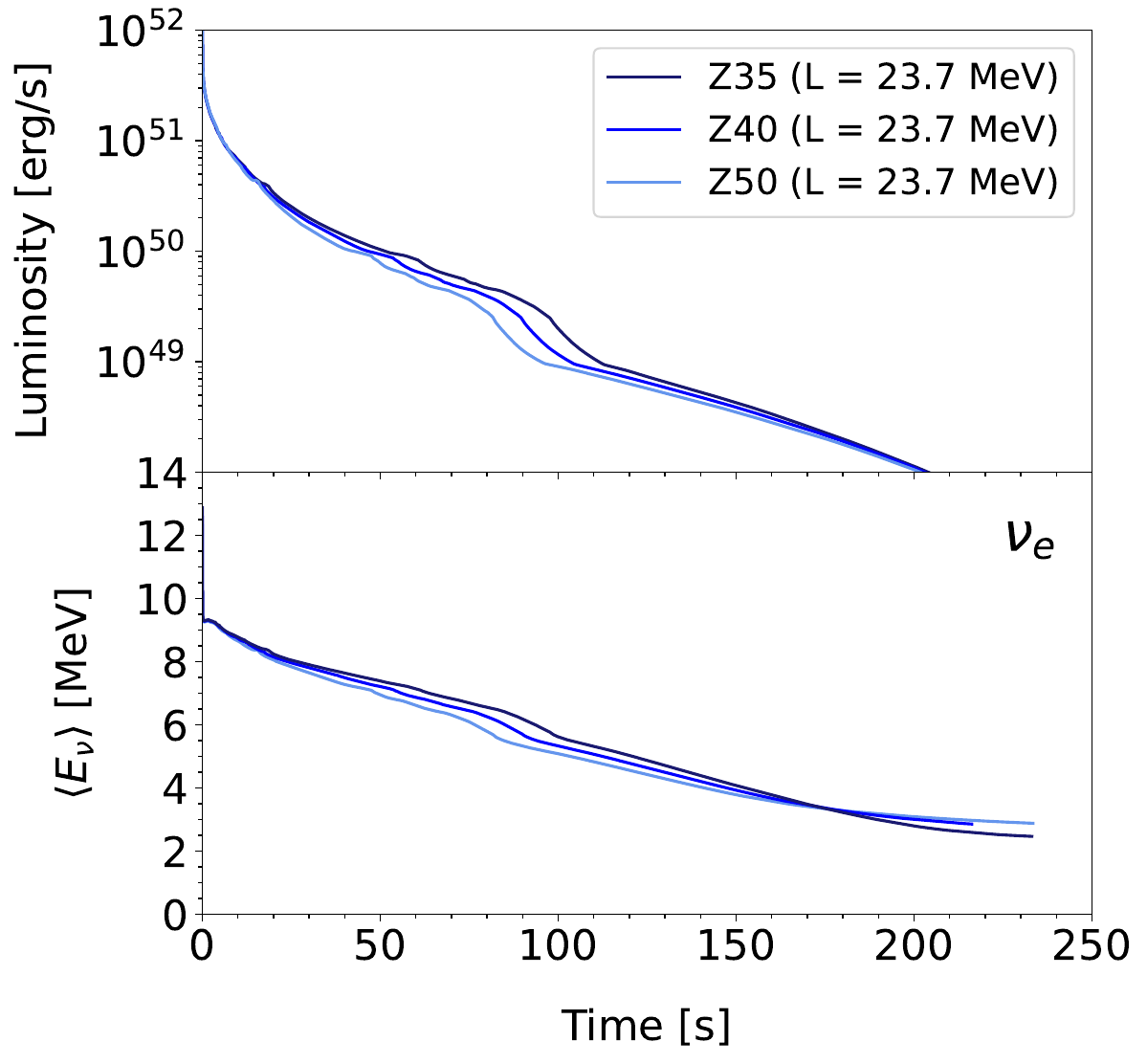}
    \vspace{0.5ex}
    {\small (a) $\nu_e$}
  \end{minipage}
  \hfill
  \begin{minipage}[t]{0.325\textwidth}
    \centering
    \includegraphics[width=\linewidth]{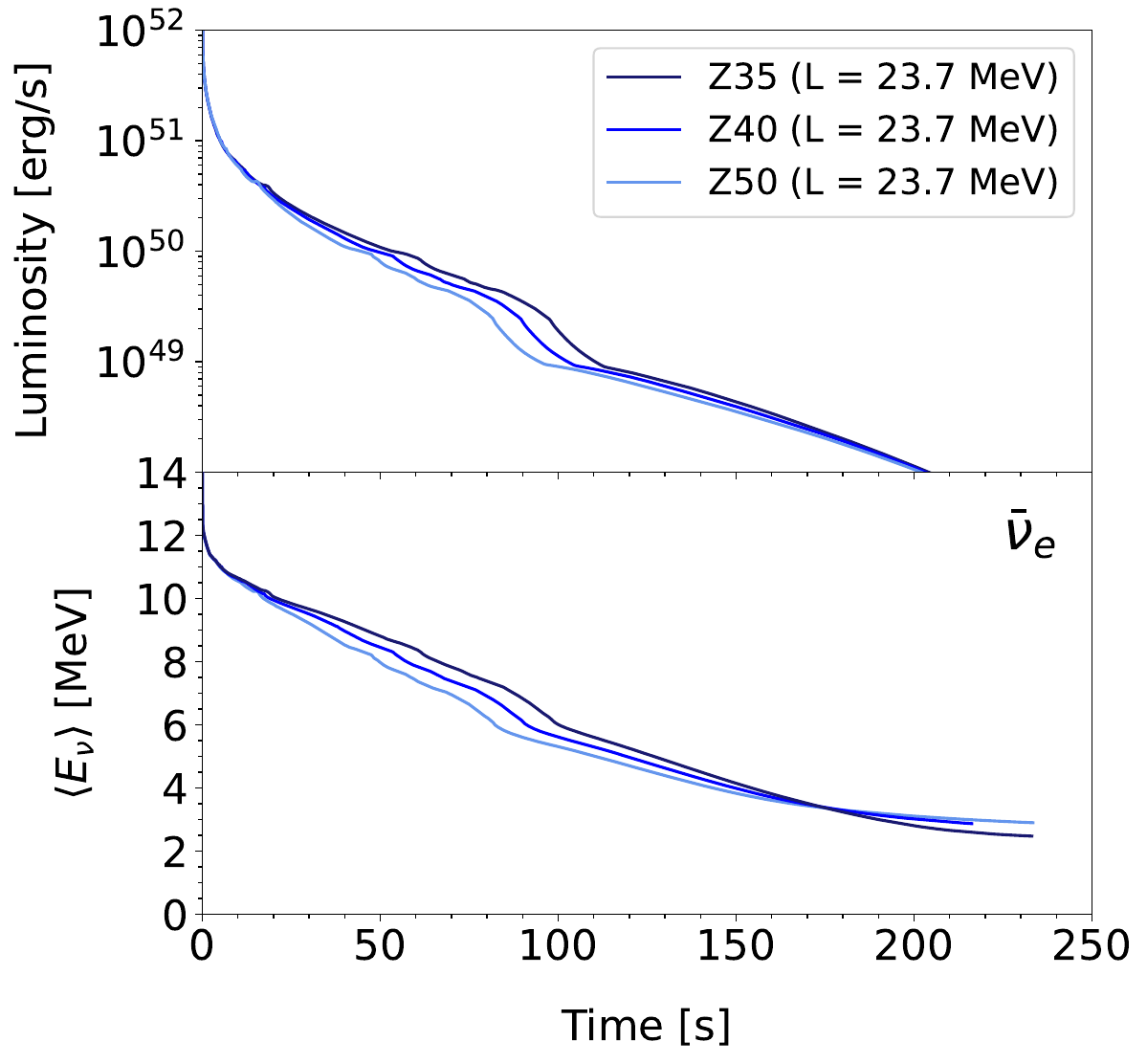}
    \vspace{0.5ex}
    {\small (b) $\bar\nu_e$}
  \end{minipage}
  \hfill
  \begin{minipage}[t]{0.325\textwidth}
    \centering
    \includegraphics[width=\linewidth]{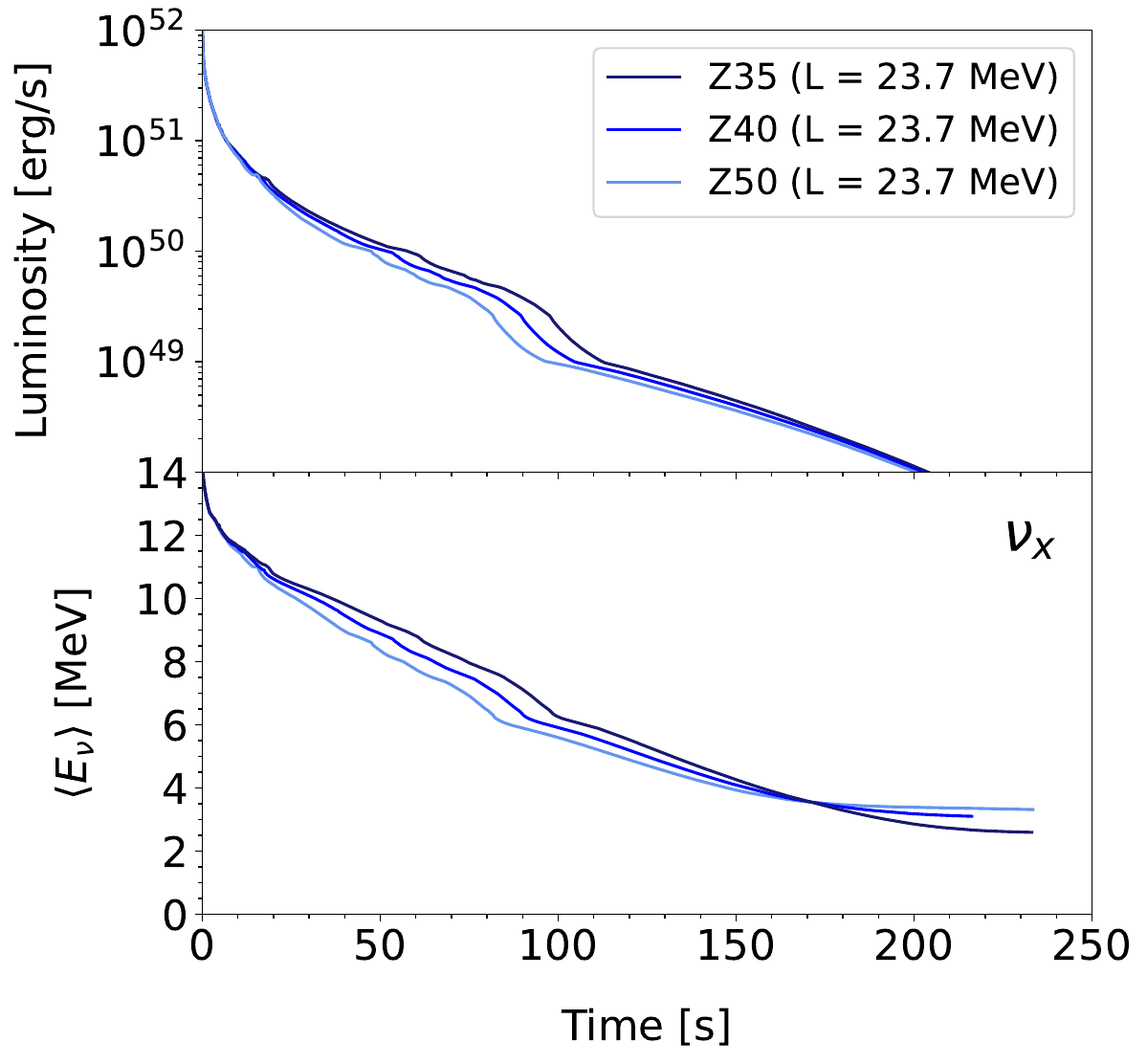}
    \vspace{0.5ex}
    {\small (c) $\nu_x$}
  \end{minipage}
\caption{Same as Fig.~\ref{fig:nulc} but for the models of EOS~Z. The line colors follow the same definitions as in Fig.~\ref{fig:MR}(c).}
\label{fig:nulcZ}
\end{figure}

In Figs.~\ref{fig:nulcE} and \ref{fig:nulcZ}, the dependence of neutrino emission on $S_{00}$ is shown. Differences due to $S_{00}$ appear even in the relatively early phase when neutrino luminosity exceeds $10^{50}$~erg~s$^{-1}$; a larger $S_{00}$ corresponds to a shorter cooling timescale. The radius of neutron stars is larger for larger values of $S_{00}$, as previously mentioned, and the cooling is more rapid for protoneutron stars with larger radii, as evaluated from the Kelvin--Helmholtz timescale \cite{2020ApJ...891..156N}. However, the impact of $S_{00}$ does not grow significantly even in the later stages when the neutrino luminosity has decreased, in contrast to the effect of $L$ on the evolution.

\begin{figure}[t]
\centering
  \begin{minipage}[t]{0.495\textwidth}
    \centering
    \includegraphics[width=\linewidth]{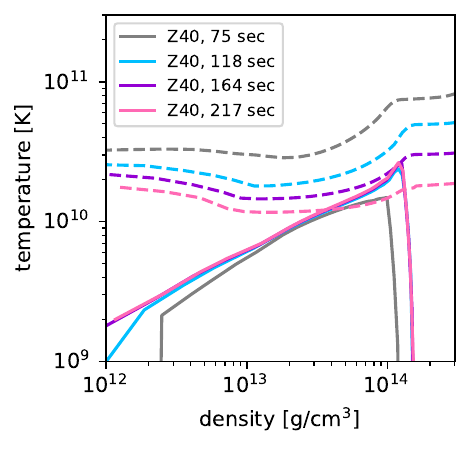}
    \vspace{0.5ex}
    {\small (a) Model Z40 ($L=23.7$~MeV)}
  \end{minipage}
  \hfill
  \begin{minipage}[t]{0.495\textwidth}
    \centering
    \includegraphics[width=\linewidth]{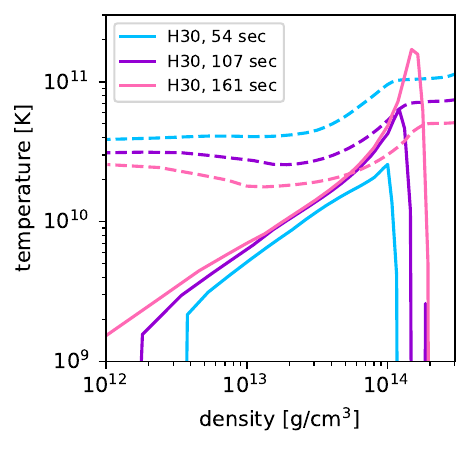}
    \vspace{0.5ex}
    {\small (b) Model H30 ($L=7.6$~MeV)}
  \end{minipage}
\caption{Time evolution of the temperature profiles for the models (a) Z40 and (b) H30. Dashed lines indicate the matter temperature of the protoneutron star as a function of density, while solid lines represent the crystallization temperature, as defined in Eq.~(\ref{eq:crit-temp}) being $\Gamma=175$.}
\label{fig:evol}
\end{figure}

We now turn to the crust formation inside the protoneutron stars. Based on the results of the protoneutron star cooling described above, we evaluate the time required for the matter temperature to decrease to the crystallization condition. For this purpose, as shown in Fig.~\ref{fig:evol}, the profiles of matter temperature and crystallization temperature are compared at each time step. The crystallization temperature, given by Eq.~(\ref{eq:crit-temp}) with $\Gamma=175$, is evaluated using the values of $Z$ and $A^\prime$ obtained from the EOS at each coordinate within the protoneutron star. In general, the matter temperature in a protoneutron star decreases over time, while the crystallization temperature increases due to the rise in $Z$ caused by cooling. We then define the crust formation time as the moment when the matter temperature first falls below the crystallization temperature at some point within the protoneutron star. For example, in the case of model Z40 ($L=23.7$~MeV), this occurs at $t=164$~s, where $t$ denotes the time measured from the onset of protoneutron star cooling. In contrast, for model H30 ($L=7.6$~MeV), crust formation occurs earlier, at $t=107$~s (see Fig.~\ref{fig:evol}), because a smaller $L$ results in a higher crystallization temperature despite slower cooling and a higher matter temperature.

\begin{figure}[htpb]
\centering
  \begin{minipage}[t]{0.495\textwidth}
    \centering
    \includegraphics[width=\linewidth]{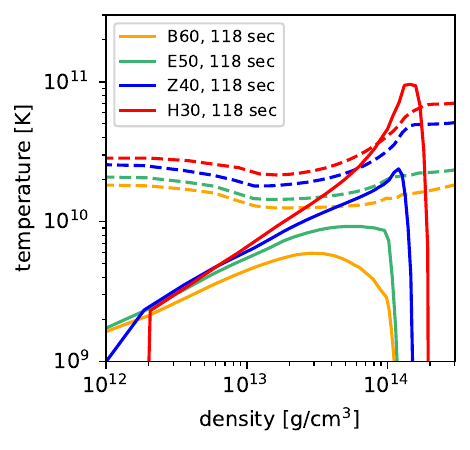}
    \vspace{0.5ex}
    {\small (a) $t=118$~s}
  \end{minipage}
  \hfill
  \begin{minipage}[t]{0.495\textwidth}
    \centering
    \includegraphics[width=\linewidth]{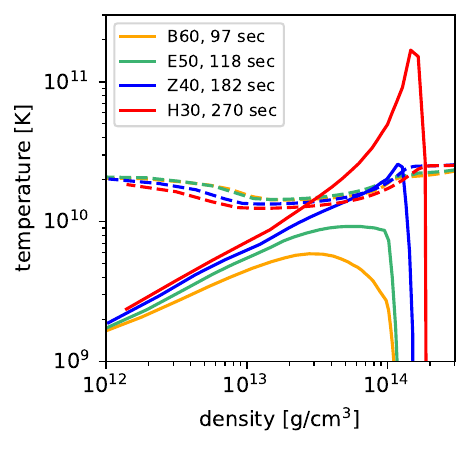}
    \vspace{0.5ex}
    {\small (b) $T_{\rm center}=3\times10^{10}$~K}
  \end{minipage}
\caption{Comparison of temperature profiles among models at (a) $t=118$~s and (b) the time when the central temperature reaches $T_{\rm center}=3\times10^{10}$~K. The line colors follow the same definitions as in Fig.~\ref{fig:MR}(a). The solid and dashed lines follow the same definitions as in Fig.~\ref{fig:evol}}
\label{fig:prof}
\end{figure}

In Fig.~\ref{fig:prof}, we focus on the $L$ dependence in the comparison between the matter temperature and the crystallization temperature for the reference models. Panel (a) shows that, when compared at the same time $t$, both the matter temperature and the crystallization temperature are higher for smaller $L$, but the latter exhibits a stronger dependence on $L$. This indicates that the $L$ parameter plays a role in crust formation primarily by determining the crystallization temperature, rather than by affecting the cooling process. Indeed, as previously noted, crust formation occurs earlier in model H30 with a lower $L$ than in model Z40. Panel (b) presents a comparison made at the moment when the central temperature of the protoneutron star reaches $T_{\rm center}=3\times10^{10}$~K, highlighting the $L$ dependence of the crystallization temperature. Unfortunately, our current protoneutron star cooling code cannot track the evolution down to significantly lower temperatures beyond this point, which prevents us from reaching the crust formation stage in models B60 and E50.

\begin{figure}[htpb]
\centering
  \begin{minipage}[t]{0.495\textwidth}
    \centering
    \includegraphics[width=\linewidth]{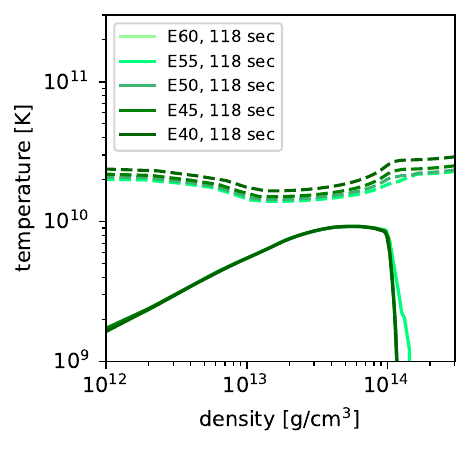}
    \vspace{0.5ex}
    {\small (a) EOS E ($L=42.6$~MeV), $t=118$~s}
  \end{minipage}
  \hfill
  \begin{minipage}[t]{0.495\textwidth}
    \centering
    \includegraphics[width=\linewidth]{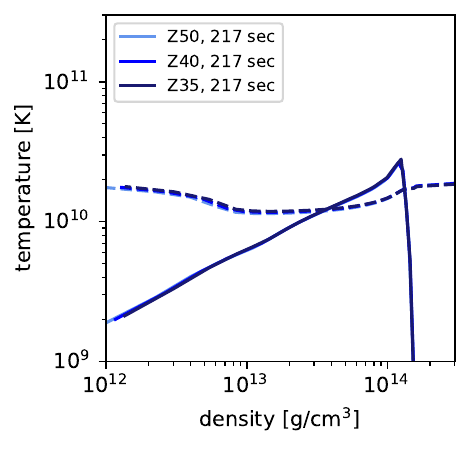}
    \vspace{0.5ex}
    {\small (b) EOS Z ($L=23.7$~MeV), $t=217$~s}
  \end{minipage}
\caption{Comparison of temperature profiles among models of (a) EOS E at $t=118$~s and (b) EOS Z at $t=217$~s. The line colors in (a) and (b) follow the same definitions as in Fig.~\ref{fig:MR}(b) and Fig.~\ref{fig:MR}(c), respectively. The solid and dashed lines follow the same definitions as in Fig.~\ref{fig:evol}.}
\label{fig:profEZ}
\end{figure}

In Fig.~\ref{fig:profEZ}, we investigate the $S_{00}$ dependence by comparing models based on EOSs E and Z, where each panel displays models with the same subnuclear EOS at a fixed time $t$. This figure demonstrates that not only the crystallization temperature but even the matter temperature shows almost no dependence on $S_{00}$. The crust formation times for models with EOS Z are 160~s for model Z50, 164~s for model Z40, and 166~s for model Z35, again indicating insensitivity to $S_{00}$. Since $S_{00}$ characterizes the EOS at supranuclear densities, it is reasonable to conclude that it does not play a major role in crust formation in the subnuclear region.

\section{Conclusions and Discussion}
\label{sec:summary}
In this paper, we have constructed a series of nuclear matter EOS that incorporate inhomogeneous phases and finite temperatures, with varying values of the symmetry energy slope $L$. For this purpose, we have used interaction parameters updated in Ref.~\cite{2023PTEP.2023f3D03O} and extended the model to finite temperatures using the free energy density of a non-interacting Fermi gas. The inhomogeneous phase has been modeled employing the Thomas--Fermi approximation, in accordance with Ref.~\cite{2023PTEP.2023f3D03O}. As a result, we have found that the crystallization temperature of the inhomogeneous phase is higher for EOS models with smaller values of $L$ because the proton numbers are larger and the Coulomb energy is enhanced. Furthermore, our EOS models have been applied to cooling simulations of protoneutron stars. To this end, the EOS has been extended to supranuclear densities. We have found that the crust formation time is earlier for EOS models with smaller values of $L$ due to the higher crystallization temperature, while it is less dependent on the EOS at supranuclear densities.

It is interesting to consider potential signatures of crust formation in a protoneutron star, as inferred from neutrino observations. During the Kelvin--Helmholtz phase, the neutrino cooling timescale depends on the neutron star radius, which depends not only on $L$ but also on the symmetry energy at supranuclear densities. In contrast, the features of neutrino emission in the later low-luminosity ($\lesssim10^{49}$~erg~s$^{-1}$) phase are governed by the mass number of nuclei in the inhomogeneous nuclear matter, which is set by $L$; the EOS models with smaller values of $L$ exhibit long-lasting neutrino emission at low luminosities. Nonetheless, it is challenging to identify a clear trend in observables such as the neutrino luminosity at the onset of crust formation because the EOS models in which neutrino emission persists for a longer period tend to undergo earlier crust formation. In principle, we would first estimate the mass and radius of the neutron star from the total energy of emitted neutrinos and the Kelvin--Helmholtz timescale, then deduce $L$ from neutrino observations during the later phase based on the mass and radius estimations, and finally use these estimates to determine the crust formation time.

Although this study has provided new insights into crust formation in protoneutron stars, the inclusion of additional processes and effects neglected here would be required for further refinement. In particular, the present simulation has not been extended sufficiently beyond the point at which the protoneutron star becomes transparent to neutrinos. This phase corresponds to a timescale from several minutes to days after the core collapse. For EOS models with larger values of $L$, crust formation is expected to occur within this timeframe. For investigating the evolution during this phase, the standard neutron star cooling framework \cite{2004ARA&A..42..169Y,2006NuPhA.777..497P} should be extended as implemented in Ref.~\cite{2020ApJ...888...97B}.

The modeling of inhomogeneous nuclear matter and the criterion for crystallization are subjects that warrant further examination. In the present work, we have employed a Thomas--Fermi approach and adopted the Coulomb coupling parameter threshold of $\Gamma=175$. While this choice corresponds to the classical melting criterion for a one-component plasma, consideration of a multi-component system could increase the threshold value and thereby lower the crystallization temperature \cite{2007PhRvE..75f6101H}. Furthermore, at subnuclear densities, nonspherical nuclear configurations, collectively referred to as nuclear pasta, can appear in equilibrium \cite{1983PhRvL..50.2066R,1984PThPh..71..320H,1993NuPhA.561..431O}. For EOS models with smaller values of $L$, a density range of pasta phases tends to be wider \cite{2007PhRvC..75a5801O}. While crystallization in multi-component systems and pasta structures at finite temperatures were investigated via molecular dynamics simulations \cite{2004PhRvC..69e5805W,2008PhRvC..77c5806S,2007PhRvE..75f6101H,2017RvMP...89d1002C}, an EOS incorporating these over an extensive range of temperatures and densities is required for application to protoneutron star cooling simulations. This remains a challenging task for future work.

\section*{Acknowledgment}
The authors thank Kazuhiro Oyamatsu, Yudai Suwa, and Akira Dohi for valuable comments. This work was supported by Grants-in-Aid for Scientific Research (JP20K03973, JP21K13924, JP24K07021, JP24K00632) and Grant-in-Aid for Transformative Research Areas (JP24H02245, JP24H01817, JP25H01273) from the Ministry of Education, Culture, Sports, Science and Technology (MEXT), Japan. In this work, numerical computations were partially performed on the supercomputers at Research Center for Nuclear Physics (RCNP) in The University of Osaka.
For providing high performance computing resources, KS acknowledges KEK, JLDG, RCNP, and YITP Kyoto University.  

\bibliographystyle{ptephy4ads}
\bibliography{bib}
%

\end{document}